\documentclass[aps,prd,twocolumn,floatfix,superscriptaddress,nofootinbib]{revtex4-2}
\usepackage[utf8]{inputenc}
\usepackage{graphicx,epsfig}
\usepackage{amsmath}
\usepackage{amssymb}
\usepackage{multirow}
\usepackage{float}
\usepackage{ulem}
\usepackage[dvipsnames]{xcolor}
\usepackage{bm,mathrsfs}
\definecolor{RevisionOrange}{RGB}{194,88,0}
\providecommand{\revisioncolor}{RevisionOrange}

\usepackage[colorlinks]{hyperref}

\hypersetup{hidelinks,pdfstartview={FitH},breaklinks=true}

\begin{document}

\title{Tidal Response Beyond the Background Geometry in Einstein-Euler-Heisenberg Black Holes}
%\title{Quadrupolar logarithmic tidal response of Einstein–Euler–Heisenberg black holes: the role of the second electromagnetic invariant\\}

\author{Ram\'on B\'ecar}
\email{rbecar@uct.cl}
\affiliation{Departamento de Ciencias Matem\'aticas y F\'isicas, Facultad de Ingenier\'ia,
Universidad Cat\'olica de Temuco, Montt 56, Casilla 15-D, Temuco, Chile}

\author{P. A. Gonz\'alez}
\email{pablo.gonzalez@udp.cl}
\affiliation{Facultad de Ingenier\'ia y Ciencias, Universidad Diego Portales,
Avenida Ej\'ercito Libertador 441, Casilla 298-V, Santiago, Chile}

\author{Ali \"Ovg\"un}
\email{ali.ovgun@emu.edu.tr}
\affiliation{Physics Department, Faculty of Arts and Sciences, Eastern Mediterranean University,
Famagusta, 99628 North Cyprus via Mersin 10, Turkiye}

\author{Joel Saavedra}
\email{joel.saavedra@pucv.cl}
\affiliation{Instituto de F\'isica, Pontificia Universidad Cat\'olica de Valpara\'iso,
Casilla 4059, Valpara\'iso, Chile}

\author{Yerko V\'asquez}
\email{yvasquez@userena.cl}
\affiliation{Departamento de F\'isica, Facultad de Ciencias, Universidad de La Serena,
Avenida Cisternas 1200, La Serena, Chile}

\begin{abstract}

We determine the quadrupolar logarithmic tidal response of
electrically charged Einstein-Euler-Heisenberg black holes from the coupled static odd-parity gravitational and electromagnetic equations. Although the second electromagnetic invariant vanishes on the electric background, its quadratic interaction contributes to the axial fluctuation equations. At first order in the
Euler-Heisenberg coupling $a$, a direct asymptotic recursion gives the canonical running matrix
$C_{\rm can}^{\rm EH}=(9aQ^2/10)\left(\begin{smallmatrix}0&-Q\\-Q&M\end{smallmatrix}\right)$, defined as the coefficients multiplying the canonical source vector in the  $r^{-2}\log(r_0/r)$ term. The resulting local coefficients retain the full dependence on the subextremal ratio $\frac{Q}{M}$ at this perturbative order. The two quartic invariants partially cancel in both nonzero channels, while the action-normalized fields make gravitoelectromagnetic reciprocity explicit. The local extremal limit of the running matrix reproduces the corresponding two-invariant effective-field-theory result after convention conversion. Allowing independent quartic couplings further shows that identical electric background geometries can have different axial running matrices and yields algebraic inverse relations within the specified two-operator model. The calculation determines local logarithmic running within minimally coupled quartic electrodynamics; the finite horizon-matched response requires an additional global matching calculation.

\end{abstract}

\maketitle
\tableofcontents

\newpage

\section{Introduction} 
\label{sec:intro} 

The tidal response of a compact object is encoded in the multipole moments induced by an external gravitational or electromagnetic field. In compact-binary dynamics, the relevant response coefficients contain finite-size information beyond the mass, spin, and conserved charges \cite{Flanagan:2007ix,Binnington:2009bb,Goldberger:2004jt}. The static Love numbers of four-dimensional Schwarzschild black holes are zero. For electrically charged black holes, gravitational and electromagnetic perturbations are coupled, but the conservative response of the Reissner-Nordstrom solution also vanishes in Einstein-Maxwell theory after growing and decaying branches of the source are separated consistently \cite{Pereniguez:2021xcj,Rai:2024lho}. This separation is nontrivial because the subleading terms of a growing solution may have the same radial behavior as that of an induced multipole.

Higher dimensional interactions can change the conclusion, producing a nonvanishing response and logarithmic mixing between the growing and decaying asymptotic branches \cite{Barbosa:2025uau}. The logarithmic terms are scale dependent, but alone do not fix the finite Love coefficients. Response matrices for charged black-holes for a tower of electromagnetic operators depending on a single field invariant have been calculated by Barbosa, Fichet and de Souza \cite{Barbosa:2026qcv}. Noumi and Wong then studied extremal charged black holes in an effective Einstein-Maxwell field theory with independent quartic electromagnetic invariants and demonstrated reciprocal mixed logarithmic responses \cite{Noumi:2026shc}. These results show that the tidal response probes information about the underlying electromagnetic effective theory which is not present in the Einstein-Maxwell background alone.

A more fundamental question is whether the response is uniquely determined by the corrected black-hole geometry. Nonlinear electrodynamics provides a sharp counterexample. Consider an electromagnetic Lagrangian depending on the two invariants $ F=F_{\mu\nu}F^{\mu\nu}/4$ and $ G=F_{\mu\nu}{}^{\star}F^{\mu\nu}/4$. In a purely electric background, \(\overline{ G}=0\), and hence the quartic interaction proportional to \( G^{2}\) will not contribute to the background field equations. Nevertheless, its second variation contains $(\delta G)^{2}$, which is generically nonzero in the odd-parity sector and contributes to the quadratic perturbation action.
Thus, two theories can produce the same electric black-hole geometry but lead to different axial response matrices.
Therefore, the tidal response can break a degeneracy between theories at the level of the background solution.
The Einstein-Euler-Heisenberg theory provides a physically motivated realization of this mechanism. Its weak-field expansion describes the leading nonlinear electromagnetic interactions generated by charged quantum fields \cite{Dunne:2004nc}. These interactions, when coupled to general relativity, give rise to charged black-hole solutions whose causal, thermodynamic, and optical properties have been studied in a variety of settings \cite{Yajima:2000kw,Ruffini:2013hia,Magos:2020ykt,Luo:2026srx,Luo:2026ndd,Breton:2021mju,Lambiase:2024lvo}. The perturbation theory of self-gravitating nonlinear
electrodynamics also demonstrates that the fluctuation
equations depend on the constitutive derivatives of the
complete electromagnetic Lagrangian. General stability
conditions were derived by Moreno and Sarbach
\cite{Moreno:2002gg}. Coupled gravitoelectromagnetic
quasinormal spectra and the breaking of the
Reissner--Nordstr\"om axial--polar isospectrality were
investigated by Chaverra et al.~\cite{Chaverra:2016ttw}.
Daghigh and Green subsequently derived the coupled odd-
and even-parity perturbation equations for electrically
charged black holes in general two-invariant nonlinear
electrodynamics \cite{Daghigh:2021psm}. In the specific
Euler--Heisenberg theory, corrections to the quasinormal
frequencies were computed including the quartic
electromagnetic invariants \cite{Nomura:2021}.

In this work we calculate the quadrupolar static odd-parity logarithmic response of electrically charged
Einstein-Euler-Heisenberg black holes for a general subextremal charge to mass ratio, to first order in the quartic electromagnetic couplings. We obtain the gravitational and electromagnetic equations in coupled form directly from the two-invariant theory and write the asymptotic response in an action-normalized basis. Direct recursion of the large-radius equations yields the complete canonical running matrix $C_{\rm can}^{\rm EH}=(9aQ^2/10)\left(\begin{smallmatrix}0&-Q\\-Q&M\end{smallmatrix}\right)$, the coefficient in the term \(r^{-2}\log(r_{0}/r)\) multipliing the canonical source vector. The off-diagonal components are equal, which expresses gravitoelectromagnetic reciprocity explicitly.
We obtain the separate contributions of the operators $ F^{2}$ and $ G^{2}$. The first-invariant contribution enables a direct comparison to the one-invariant calculation of Ref.~\cite{Barbosa:2026qcv} while the second-invariant contribution generalizes the corresponding extremal result of Ref.~\cite{Noumi:2026shc} to generic subextremal $M$ and $Q$. In the nonzero channels the two operators partially cancel, indicating that the second invariant cannot be dropped from the Euler-Heisenberg response even when it is not visible in the electric background geometry. We also allow the two quartic couplings to vary independently and find the algebraic relation to the entries of the running matrix. In this two-operator model, the axial response thus lifts a degeneracy that is not lifted with the electric background alone.
As an independent check, we numerically integrate the full static perturbation equations, impose the regularity conditions at the outer horizon, and extract the asymptotic source and logarithmic-response matrices. The term linear in the quartic coupling is extracted numerically by a centered finite difference, and the resulting matrices for representative subextremal charges reproduce the analytic running coefficients. The numerical calculation also recovers the vanishing diagonal entry in the gravitational and the reciprocity relation between the mixed components. Thus, it allows a direct test of the local logarithmic coefficient without recourse to the asymptotic recursion used in its analytic derivation.
The scope of the result must be distinguished from a calculation of the complete finite tidal response. The resonant asymptotic recursion sets the logarithmic coefficient locally, whereas the finite part is set by a global horizon-regular solution and a given matching and subtraction prescription. The finite extremal limit of the local running matrix does not in itself show a uniform extremal limit for the horizon-matched response.
Furthermore, here the minimally coupled quartic electromagnetic action is a constrained effective theory. The complete low-energy quantum-electrodynamical action in a curved spacetime also contains curvature-electromagnetic and derivative operators \cite{Bastianelli:2008}. We isolate in our result the contribution of the two Euler-Heisenberg quartic invariants, and do not claim to contain all the curved-spacetime QED corrections, or to provide a complete binary-waveform prediction.

The work is organized as follows. Section~\ref{sec:background} specifies the theory and its validity domain. Section~\ref{sec:axial} gives the static equations; Sec.~\ref{sec:response} derives and interprets the running matrices. Section~\ref{sec:degeneracy} develops the independent-coupling extension. Finally, Section~\ref{sec:conclusions} contains our conclusions. The appendices~\ref{app:axial_derivation}
and~\ref{app:asymptotic-recursion} provide canonical normalization and resonant recursion. 

\section{Einstein-Euler-Heisenberg theory and black-hole background}
\label{sec:background}

We consider four-dimensional general relativity minimally coupled to Euler-Heisenberg nonlinear electrodynamics. The effective Euler-Heisenberg theory provides the leading nonlinear quantum-electrodynamical corrections to Maxwell theory in the low-energy and weak-field regime
\cite{Euler:1935zz,Heisenberg:1936nmg,Dunne:2004nc}.
Throughout this work, we adopt the metric signature $(-,+,+,+)$ and define the two independent electromagnetic invariants as
\begin{equation}
    F \equiv \frac{1}{4}F_{\mu\nu}F^{\mu\nu},
    \qquad
    G \equiv \frac{1}{4}F_{\mu\nu}{}^{\star}F^{\mu\nu},
    \label{eq:FG_def}
\end{equation}
where
\begin{equation}
    {}^{\star}F^{\mu\nu}
    =
    \frac{1}{2}\epsilon^{\mu\nu\rho\sigma}F_{\rho\sigma}
\end{equation}
denotes the dual electromagnetic tensor. With these conventions, the action takes the form
\begin{equation}
    S=
    \int d^4x\,\sqrt{-g}
    \left[
        \frac{R}{16\pi}
        -\frac{L(F,G)}{4\pi}
    \right],
    \label{eq:action_EEH}
\end{equation}
where $R$ is the Ricci scalar. At leading nonlinear order, the Euler-Heisenberg theory is described by the Euler-Kockel
truncation
\begin{equation}
    L(F,G)
    =
    F
    -\frac{a}{2}F^2
    -\frac{7a}{8}G^2
    +\mathcal{O}(a^2) ,
    \label{eq:EH_lagrangian}
\end{equation}
where $a$ denotes the Euler-Heisenberg coupling. This convention is equivalent to those commonly adopted in the literature after taking into account the different normalizations of electromagnetic invariants \cite{Dunne:2004nc,Breton:2021mju}. The Maxwell theory is
recovered in the limit $a\rightarrow0$. Since Eq.~\eqref{eq:EH_lagrangian} corresponds to a weak-field expansion, all background and perturbative quantities considered below are consistently retained only through first order in $a$.

For a general NLED theory depending on both electromagnetic
invariants; the variation of Eq.~\eqref{eq:action_EEH} with respect to the electromagnetic potential yields
\begin{equation}
    \nabla_\mu
    \left(
        L_F F^{\mu\nu}
        +L_G\,{}^{\star}F^{\mu\nu}
    \right)=0,
    \label{eq:NLED_Maxwell}
\end{equation}
where
\begin{equation}
    L_F\equiv\frac{\partial L}{\partial F},
    \qquad
    L_G\equiv\frac{\partial L}{\partial G},
\end{equation}
together with the Bianchi identity
\begin{equation}
    \nabla_\mu {}^{\star}F^{\mu\nu}=0.
\end{equation}
Variation with respect to the metric gives
\begin{equation}
    G_{\mu\nu}=8\pi T_{\mu\nu},
    \label{eq:Einstein_EEH}
\end{equation}
with
\begin{equation}
    T_{\mu\nu}
    =
    \frac{1}{4\pi}
    \left[
        L_F F_{\mu\rho}F_{\nu}{}^{\rho}
        +g_{\mu\nu}\left(L_GG-L\right)
    \right].
    \label{eq:Tmunu_NLED}
\end{equation}
These equations provide the starting point for both the background solution and the coupled gravitational-electromagnetic perturbation analysis. In particular, in self-gravitating NLED theories, the perturbation equations depend on derivatives of the complete electromagnetic Lagrangian and cannot, in general, be reconstructed from the background metric alone
\cite{Moreno:2002gg,Daghigh:2021psm}.

We focus on the electrically charged, static, and spherically symmetric Einstein-Euler-Heisenberg black-hole solution
\cite{Yajima:2000kw,Breton:2021mju}.
The line element can be written as
\begin{equation}
    ds^2
    =
    -f(r)\,dt^2
    +\frac{dr^2}{f(r)}
    +r^2
    \left(
        d\theta^2+\sin^2\theta\,d\phi^2
    \right),
    \label{eq:EEH_metric}
\end{equation}
and the background electromagnetic field is purely electric,
\begin{equation}
    F_{tr}=E(r).
    \label{eq:electric_ansatz}
\end{equation}
For such a configuration, the pseudoscalar electromagnetic invariant vanishes identically,
\begin{equation}
    \bar G=0,
    \label{eq:G_background}
\end{equation}
whereas
\begin{equation}
    \bar F=-\frac{E^2(r)}{2}.
    \label{eq:F_background}
\end{equation}
Consequently, $L_G=0$ on the background and
Eq.~\eqref{eq:NLED_Maxwell} reduces to a first integral,
\begin{equation}
    r^2 L_F E(r)=Q,
    \label{eq:electric_first_integral}
\end{equation}
where $Q$ is identified with the conserved electric charge.

For the Euler-Heisenberg Lagrangian \eqref{eq:EH_lagrangian}, the relevant constitutive derivatives are
\begin{equation}
    L_F=1-aF,
    \qquad
    L_G=-\frac{7a}{4}G,
    \qquad
    L_{GG}=-\frac{7a}{4}.
    \label{eq:EH_derivatives}
\end{equation}
Solving Eq.~\eqref{eq:electric_first_integral} perturbatively to
$\mathcal{O}(a)$ gives
\begin{equation}
    E(r)
    =
    F_{tr}
    =
    \frac{Q}{r^2}
    \left(
        1-\frac{aQ^2}{2r^4}
    \right)
    +\mathcal{O}(a^2),
    \label{eq:electric_field}
\end{equation}
and therefore
\begin{equation}
     F
    =
    -\frac{Q^2}{2r^4}
    +\mathcal{O}(a),
    \qquad
    L_F
    =
    1+\frac{aQ^2}{2r^4}
    +\mathcal{O}(a^2).
    \label{eq:background_invariants}
\end{equation}
This electric-field deformation represents the leading nonlinear vacuum-polarization correction to the Coulomb field \cite{Breton:2021mju}.

It is useful to introduce the mass function $m(r)$ according to
\begin{equation}
    f(r)=1-\frac{2m(r)}{r}.
    \label{eq:mass_function_def}
\end{equation}
Solving the Einstein equations consistently through first order in $a$ yields
\begin{equation}
    m(r)
    =
    M
    -\frac{Q^2}{2r}
    +\frac{aQ^4}{40r^5}
    +\mathcal{O}(a^2),
    \label{eq:mass_function}
\end{equation}
where $M$ is the ADM mass. Hence, the metric function is
\begin{equation}
    f(r)
    =
    1-\frac{2M}{r}
    +\frac{Q^2}{r^2}
    -\frac{aQ^4}{20r^6}
    +\mathcal{O}(a^2)\,.
    \label{eq:EEH_f}
\end{equation}
This form of the electrically charged EEH solution agrees, after translating between the different conventions for the coupling and electromagnetic invariants, with the black-hole geometries obtained in Refs.~\cite{Yajima:2000kw,Breton:2021mju}.
In particular, Eq.~\eqref{eq:EEH_f} may alternatively be written as
\begin{equation}
    f(r)
    =
    1-\frac{2M}{r}
    +\frac{Q^2}{r^2}
    \left(
        1-\frac{aQ^2}{20r^4}
    \right)
    +\mathcal{O}(a^2),
    \label{eq:EEH_f_screening}
\end{equation}
which makes explicit the leading Euler-Heisenberg correction to the Reissner-Nordstr\"om charge contribution. This correction can be interpreted as a screening of the electric charge associated with vacuum-polarization effects
\cite{Breton:2021mju}.

The event horizons are determined by the positive roots of
$f(r)=0$. In particular, the outer event horizon $r_+$ is the largest
positive root of
\begin{equation}
    r_+^6
    -2Mr_+^5
    +Q^2r_+^4
    -\frac{aQ^4}{20}
    =0.
    \label{eq:horizon_EEH}
\end{equation}
Depending on the values of $Q$ and the Euler-Heisenberg coupling, the horizon structure can differ from that of the
Reissner-Nordstr\"om geometry \cite{Yajima:2000kw,Breton:2021mju}. Extremal configurations are characterized by the simultaneous
conditions
\begin{equation}
    f(r_{\rm e})=0,
    \qquad
    f'(r_{\rm e})=0.
    \label{eq:extremality_EEH}
\end{equation}
The Einstein-Maxwell limit is recovered for $a\rightarrow0$,
\begin{equation}
    f(r)
    \longrightarrow
    1-\frac{2M}{r}+\frac{Q^2}{r^2},
\end{equation}
corresponding to the Reissner-Nordstr\"om black hole, while the additional neutral limit $Q\rightarrow0$ yields the Schwarzschild geometry.

At a large radius,
\begin{equation}
    f(r)
    =
    1-\frac{2M}{r}
    +\frac{Q^2}{r^2}
    +\mathcal{O}(r^{-6}),
    \label{eq:asymptotic_background}
\end{equation}
so that the Euler-Heisenberg correction does not alter the standard asymptotic identification of the ADM mass and electric charge. This asymptotically flat structure will be particularly useful in the tidal-response problem, since the external tidal sources and the induced multipole moments are identified from the large-$r$ behavior of the static perturbations.

Finally, an important distinction must be emphasized before turning to the perturbation problem. Although the electrically charged background satisfies
\begin{equation}
    \bar G=0,
\end{equation}
and therefore, the $G^2$ interaction in
Eq.~\eqref{eq:EH_lagrangian} does not contribute to the background metric \eqref{eq:EEH_f}, this term is not dynamically irrelevant. Indeed,
\begin{equation}
    \left.
    \delta^2(G^2)
    \right|_{\bar G=0}
    =
    2(\delta G)^2,
    \label{eq:G2_second_variation}
\end{equation}
which is generically nonzero in the odd-parity electromagnetic sector. Thus, the absence of the second invariant from the static geometry must not be confused with its absence from the fluctuation problem. As we show in the following section, the nonvanishing quantity $L_{GG}$ modifies the axial electromagnetic operator and, through the coupling between the electromagnetic and gravitational perturbations, contributes directly to the gravitoelectromagnetic tidal response.

\subsection{Domain of the perturbative background}
\label{sec:validity}
We use geometrized units, so $[M]=[Q]=[r]={\rm length}$ and $[a]={\rm length}^2$. Let
\begin{equation}
r_\pm^{(0)}=M\pm\sqrt{M^2-Q^2},\qquad
\Delta=r_+^{(0)}-r_-^{(0)}.
\end{equation}
The horizon used in the perturbative tidal problem is the branch that is continuously connected to $r_+^{(0)}$ for $|Q|<M$. Expanding $f(r_+)=0$ gives
\begin{equation}
\delta r_+=\frac{aQ^4}{20[r_+^{(0)}]^4\Delta},\qquad
r_+=r_+^{(0)}+\delta r_++O(a^2).
\label{eq:horizon_shift}
\end{equation}
In addition to requiring weak nonlinearities throughout the exterior,
\begin{equation}
\epsilon_F=\frac{|a|Q^2}{[r_+^{(0)}]^4}\ll1,
\label{eq:weak_field_control}
\end{equation}
a sufficient condition for an ordinary expansion about a simple RN root is
\begin{equation}
\epsilon_{\rm hor}=\frac{|\delta r_+|}{\Delta}
=\frac{|a|Q^4}{20[r_+^{(0)}]^4\Delta^2}\ll1.
\label{eq:horizon_control}
\end{equation}
The second condition is not uniform as $|Q|\to M$. A finite limit of the asymptotic logarithmic coefficient at $Q=M$ does not establish a uniform limit of a horizon-matched solution. Extra small-radius roots of the truncated polynomial~\eqref{eq:horizon_EEH} are not reliable predictions when the weak-field expansion fails there.

For literal QED, the low-energy derivative expansion and subcritical local electric field must also be checked after restoring the charged-particle mass and electromagnetic normalization. Small $\epsilon_F$ alone does not impose these independent conditions. Curvature-electromagnetic terms such as $RF_{\mu\nu}F^{\mu\nu}$, $R_{\mu\nu}F^{\mu\rho}F^\nu{}_{\rho}$, and $R_{\mu\nu\rho\sigma}F^{\mu\nu}F^{\rho\sigma}$ are allowed in curved spacetime \cite{Bastianelli:2008}. Accordingly, the term ``complete Euler-Heisenberg matrix'' below refers to complete within the minimally coupled quartic electromagnetic action displayed.

The calculation is also restricted to electrovacuum.
Adding a perfect-fluid-dark-matter contribution to the
background lapse would require a corresponding matter
perturbation model and suitable outer boundary conditions.
In the axial sector, the relevant additional variables
include velocity and axial stress perturbations.
Therefore, environmental tidal responses cannot be inferred
from a modified lapse alone
\cite{Cardoso:2019upw,Chakraborty:2024gcr}.

\section{Odd-parity static perturbations}
\label{sec:axial}

In Regge-Wheeler gauge we expand
\begin{equation}
\begin{aligned}
\delta g_{tA}&=h_0(r)S_A^{\ell m}\,,\qquad
\delta g_{rA}=h_1(r)S_A^{\ell m}\,,\\
\delta A_A&=u(r)S_A^{\ell m}\,,
\end{aligned}
\label{eq:oddpert}
\end{equation}
where $S_A^{\ell m}=\epsilon_A{}^B D_BY_{\ell m}$ and $L=\ell(\ell+1)$. We restrict to $\ell\ge2$. The static angular Einstein equation gives $(fh_1)'=0$, while the independent $rA$ equation, after the use of the background equations, imposes
\begin{equation}
(L-2)h_1=0.
\label{eq:constraint}
\end{equation}
Thus, $h_1=0$ in this tidal sector. This conclusion follows from the field equations, rather than from a coordinate amplitude alone. With $dt=dv-dr/f$, regularity of the remaining perturbation in ingoing coordinates requires $h_0/f$ and $u$ to remain finite at a simple horizon.

The remaining equations, obtained from the linearized Einstein and nonlinear Maxwell equations before introducing frequency-dependent master variables, are

%\small{
\begin{align}
h_0''
-\frac{1}{f}
\left(
\frac{L}{r^2}
-\frac{4M}{r^3}
+\frac{2Q^2}{r^4}
-\frac{aQ^4}{10r^8}
\right)h_0
&=
\frac{4Q}{r^2}u',
\label{eq:static-gravitational}
\\[1ex]
\left(\mathcal{L}_{F} f u'\right)'
-\frac{L}{r^6}
\left(
r^4\mathcal{L}_{F}
+Q^2\mathcal{L}_{GG}
\right)u
&=
\frac{Q}{r^2}
\left(
h_0'-\frac{2h_0}{r}
\right).
\label{eq:static-electromagnetic}
\end{align}

The terms beyond the angular barrier in Eq.~\eqref{eq:static-gravitational} arise from the background curvature.  Defining the mass function by
\begin{equation}
m(r)\equiv\frac r2[1-f(r)]
=M-\frac{Q^2}{2r}+\frac{aQ^4}{40r^5}+\mathcal O(a^2),
\label{eq:massfunction}
\end{equation}
the gravitational potential is equivalently $L/r^2-4m(r)/r^3$.  At $a=0$ the system reduces to the standard static axial Einstein-Maxwell equations on Reissner-Nordstr\"om, after translating electromagnetic conventions.  A derivation of the static axial projections, with particular emphasis on the $ L_{ GG}$ contribution, is given in Appendix~\ref{app:axial_derivation}.

A decisive check is the Schwarzschild limit.  Setting $Q=a=0$ and $\ell=2$, Eq.~\eqref{eq:static-gravitational} becomes
\begin{equation}
h_0''-\frac1f\left(\frac6{r^2}-\frac{4M}{r^3}\right)h_0=0,
\qquad f=1-\frac{2M}{r},
\label{eq:Schwcheckeq}
\end{equation}
and its horizon-regular growing solution is exactly
\begin{equation}
h_0^{\rm Schw}=r^3-2Mr^2=r^3f.
\label{eq:Schwsolution}
\end{equation}
It does not contain a decaying response $r^{-2}$.  Retaining only $L/(fr^2)$ fails this Schwarzschild check.

For the full Euler-Heisenberg theory,
\begin{equation}
r^4 L_{F}+Q^2 L_{ GG}
=r^4-\frac{5aQ^2}{4}+\mathcal O(a^2).
\label{eq:EHoperator}
\end{equation}
By contrast, discarding $ G^2$ gives $r^4+aQ^2/2$. The discrepancy is not a higher-order correction: it occurs at the same $\mathcal O(a)$ as the deformation of the background.
The origin of this correction follows from Eq.~\eqref{eq:G2_second_variation}. The axial electromagnetic perturbation produces a nonvanishing $\delta G$, whereas $\delta G$ vanishes in the polar sector on the same electric background. The $G^2$ interaction therefore generates a parity-selective constitutive correction at the perturbative order considered here.

\section{Response matrix and logarithmic running}\label{sec:response}

At a simple outer horizon, Eqs.~\eqref{eq:static-gravitational}
and~\eqref{eq:static-electromagnetic} admit two independent
regular solutions. We define their response in the
gauge-invariant, action-normalized basis
$\boldsymbol{\Psi}=(\Psi_h,\Psi_a)^T$ introduced in
Appendix~\ref{app:axial_derivation}. The quadratic action
fixes the relative normalization of the gravitational and
electromagnetic channels and has a symmetric potential matrix.

The asymptotic source and finite response coefficients of
these solutions define the matrices $\mathbf S_\ell$ and
$\mathbf R_\ell$. For an invertible source matrix and a
specified subtraction prescription, the finite response
matrix is
\begin{equation}
\boldsymbol{\Lambda}_\ell
=
\mathbf R_\ell\mathbf S_\ell^{-1}.
\label{eq:responsematrix}
\end{equation}
All source descendants must be included before a response
coefficient is identified. In a logarithmic channel,
$\mathbf R_\ell$ denotes the finite decay coefficient
after subtraction of the logarithm on a fixed reference
scale. We introduce $\boldsymbol{\Lambda}_\ell$ to specify
the matching framework; the calculation below determines
its logarithmic running, not its finite horizon-matched value.

For integer $\ell$, the two indicial branches differ by $2\ell+1$. The $\mathcal O(a)$ inhomogeneous problem can therefore develop a resonant logarithm. The asymptotic form is schematically
\begin{equation}
\bm\Psi_\ell=r^{\ell+1}\mathbf A_\ell+\cdots
+\frac{\mathbf B_\ell(r_0)+C_\ell\mathbf A_\ell\ln(r_0/r)}{r^\ell}+\cdots.
\label{eq:asymptotic}
\end{equation}
Here $r_{0}$ is an arbitrary reference length, equivalently associated with the inverse subtraction scale $\mu=1/r_{0}$, introduced to make the argument of a logarithm dimensionless. For nonzero $C_\ell$, the finite coefficient depends on the reference scale and the subtraction prescription. The leading logarithmic running is invariant under a change of subtraction scale on a fixed canonical basis; its matrix entries still depend on field normalization and logarithm convention \cite{Barbosa:2025uau,Barbosa:2026qcv}. The complete matched response contains both finite and logarithmic terms. In this order, the logarithmic coefficient is fixed by the
asymptotic recursion of the exterior field equations without
using the horizon boundary condition. It therefore
characterizes local source-response mixing in the specified
exterior theory, rather than the finite response selected
by horizon regularity. The latter requires a global
matching problem.

\subsection{Exact quadrupolar logarithm from the second invariant}

We now isolate the effect of $\gamma\equiv\ L_{ GG}$ while holding the RN background and $ L_{ F}=1$ fixed. This separates the new operator from all $ F^2$ effects. Let the two independent raw source amplitudes be defined by
\begin{equation}
h_0^{(0)}=Hr^3+\mathcal O(r^2),\qquad
u^{(0)}=Ur^3+\mathcal O(r^2).
\label{eq:rawsources}
\end{equation}
Expanding the RN solution recursively through the resonant order and solving the first-order inhomogeneous equations give the perturbations $\delta h_0=\gamma h_0^{(1)}$ and $\delta u=\gamma u^{(1)}$:
%\small
\begin{equation}
\begin{pmatrix}\delta h_0\\\delta u\end{pmatrix}_{\log}
=\frac{\gamma\log(r_0/r)}{r^2}
\begin{pmatrix}0&6Q^3/5\\-3Q^3/10&-18MQ^2/5\end{pmatrix}
\begin{pmatrix}H\\U\end{pmatrix}.
\label{eq:rawlog}
\end{equation}

The fixed scale $r_0$ has length dimensions. Changing $r_0$ shifts only the finite response coefficients. Eq.~\eqref{eq:rawlog} is therefore independent of horizon matching and determines the local logarithmic mixing in this field basis.

For the Euler-Heisenberg value $\gamma=-7a/4$, the
second-invariant contribution in the raw field basis is
\begin{equation}
\Delta C_{\rm raw}^{(G^2),{\rm EH}}
=
a
\begin{pmatrix}
0 & -21Q^3/10\\[1mm]
21Q^3/40 & 63MQ^2/10
\end{pmatrix}.
\label{eq:rawEH}
\end{equation}
This matrix multiplies
$r^{-2}\log(r_0/r)(H,U)^T$ and includes only the
$G^2$ contribution. Its sign reverses when the logarithm
is written as $\log(r/r_0)$.

For the full static gravitational operator, the
horizon-regular quadrupolar Reissner-Nordstr\"om solution
with arbitrary growing source amplitudes $H$ and $U$
can be written exactly as
\begin{align}
h_0^{(0)}={}&Hr^3-(2MH+3QU)r^2+Q(QH+6MU)r \nonumber\\
&-2UQ^3\left(1+\frac Mr-\frac{Q^2}{2r^2}\right),
\label{eq:h0rec}
\end{align}
\begin{align}
u^{(0)}={}&Ur^3-\frac14(QH+6MU)r^2
+\frac{Q^2}{4}(QH+6MU) \nonumber\\
&-\frac{UQ^4}{r}.
\label{eq:urec}
\end{align}
Direct substitution gives zero in both RN equations.
The equivalent factorized expression
\begin{equation}
h_0^{(0)}
=
f_0(r)\left(Hr^3-3QUr^2+Q^3U\right),
\label{eq:RN_seed_factorized}
\end{equation}
where $f_0(r)=1-2M/r+Q^2/r^2$ makes the regularity
of $h_0^{(0)}/f_0$ on a simple outer horizon explicit.
The electromagnetic amplitude $u^{(0)}$ is also finite
there. The $r^{-2}$ term in the expanded metric
perturbation belongs to this source-connected RN
solution and is not an independent Love coefficient.
For $Q=U=0$, the solution reduces to
Eq.~\eqref{eq:Schwsolution}.

Writing $h_0=h_0^{(0)}+\gamma h_0^{(1)}$ and $u=u^{(0)}+\gamma u^{(1)}$, the leading forced term is $u^{(1)}=-(3/2)Q^2U/r+\cdots$. The resonant coefficients are
\begin{align}
\left.u^{(1)}\right|_{\log}
&=\left(-\frac{3Q^3}{10}H-\frac{18MQ^2}{5}U\right)
\frac{\log(r_0/r)}{r^2},\label{eq:ucorrectedlogstep}\\
\left.h_0^{(1)}\right|_{\log}
&=\frac{6Q^3}{5}U\frac{\log(r_0/r)}{r^2}.
\label{eq:hcorrectedlogstep}
\end{align}
Appendix~\ref{app:asymptotic-recursion} derives these coefficients. They are local asymptotic properties.

The apparent asymmetry of Eq.~\eqref{eq:rawEH} is not a violation of reciprocity. The action-normalized static master fields are related to the raw amplitudes by
\begin{equation}
h_0=-\alpha f[r\Psi_h]',\qquad u=\alpha\Psi_a,
\label{eq:canonicalfields}
\end{equation}
where the common nonzero factor $\alpha$ depends on the normalization of the Einstein and Maxwell terms and cancels from the response matrix. For $\ell=2$, write
\begin{equation}
\Psi_I
=
A_I r^3+\cdots+
\frac{B_I(r_0)+\mathcal K_I\log(r_0/r)}{r^2}
+\cdots,
\label{eq:masterasy}
\end{equation}
where $I=h,a$, $B_I(r_0)$ is the finite decaying
coefficient, and $\mathcal K_I$ is the logarithmic
amplitude. Equation~\eqref{eq:canonicalfields} gives
\begin{equation}
\begin{aligned}
H&=-4\alpha A_h,
&\qquad U&=\alpha A_a,\\
b_h&=\alpha\mathcal K_h,
&\qquad b_u&=\alpha\mathcal K_a.
\end{aligned}
\label{eq:normalizationmap}
\end{equation}
The factor $-4$ acts on the growing gravitational branch,
whereas the leading decaying logarithm has the factor
$+1$. The resulting canonical logarithmic amplitudes
satisfy
\begin{equation}
\begin{pmatrix}
\mathcal K_h\\
\mathcal K_a
\end{pmatrix}
=
\gamma
\begin{pmatrix}
0 & 6Q^3/5\\[1mm]
6Q^3/5 & -18MQ^2/5
\end{pmatrix}
\begin{pmatrix}
A_h\\
A_a
\end{pmatrix}.
\label{eq:canongamma}
\end{equation}
This explicitly verifies gravitoelectromagnetic reciprocity. For Euler-Heisenberg, the new contribution generated by the second electromagnetic invariant is
\begin{equation}
\Delta C^{(G^2)}_{\rm can}=a
\begin{pmatrix}
0 & -21Q^3/10\\[1mm]
-21Q^3/10 & 63MQ^2/10
\end{pmatrix}
\label{eq:canonEH}
\end{equation}
for the $\log(r_0/r)$ convention. The matrix changes sign for $\log(r/r_0)$.

The result is gauge invariant within the static odd sector. Under an odd diffeomorphism generated by $\xi_A=\xi(r)S_A$, $h_0$ changes by $\partial_t\xi$ and is therefore invariant at zero frequency; $u$ is invariant under electromagnetic gauge transformations because an axial vector harmonic cannot be generated by the gradient of a scalar gauge parameter. The local master-field definition in Appendix~\ref{app:axial_derivation} removes the integration ambiguity in reconstructing $\Psi_h$ and establishes its gauge invariance. The same prefactors in Eq.~\eqref{eq:canonicalfields} specialize to $\ell=2$.

\subsection{Complete Euler-Heisenberg running from the same recursion}
\label{sec:complete_EH_running}
At first order, the $F^2$ and $G^2$ contributions add. To calculate the first-invariant part directly, set $L_{GG}=0$ and write
\begin{equation}
f=f_0+a f_a,\quad f_a=-\frac{Q^4}{20r^6},\quad
L_F=1+a k,\quad k=\frac{Q^2}{2r^4}.
\end{equation}
Here $f_0$ is the RN lapse. Define the RN operators
\begin{align}
\mathscr D_h[h,u]&=h''-\frac{V_0}{f_0}h-\frac{4Q}{r^2}u',\\
\mathscr D_u[h,u]&=(f_0u')'-\frac{6u}{r^2}
-\frac{Q}{r^2}(h'-2h/r),\\
V_0&=\frac6{r^2}-\frac{4M}{r^3}+\frac{2Q^2}{r^4}.
\end{align}
For $h=h^{(0)}+a h_a$, $u=u^{(0)}+a u_a$, the first-order forcing is
\begin{align}
\mathscr D_h[h_a,u_a]&=\frac{Q^4}{5r^8f_0^2}h^{(0)},\label{eq:Fsourceh}\\
\mathscr D_u[h_a,u_a]&=-[(f_a+kf_0)u^{(0)\prime}]'
+\frac{6k}{r^2}u^{(0)}.\label{eq:Fsourceu}
\end{align}
The right-hand side of Eq.~\eqref{eq:Fsourceh} begins at $r^{-5}$ and does not directly force the $r^{-4}$ gravitational resonance. In the electromagnetic equation,
\begin{equation}
\mathscr D_u[h_a,u_a]=\frac{6Q^2U}{r^3}
-\frac{\frac32Q^3H+18MQ^2U}{r^4}+O(r^{-5}).
\end{equation}
With $u_a=p/r+b_u r^{-2}\log(r_0/r)+\cdots$ and $h_a=b_h r^{-2}\log(r_0/r)+\cdots$, the resonance gives
\begin{align}
p&=-\frac32Q^2U, & b_h&=\frac65Q^3U,\\
b_u&=-\frac3{10}Q^3H-\frac{27}{5}MQ^2U.
\end{align}
For the first-invariant deformation, the full electromagnetic
normalization is
\[
u=\frac{\alpha}{\sqrt{L_F}}\Psi_a
=
\alpha\left(1-\frac{aQ^2}{4r^4}\right)\Psi_a
+\mathcal O(a^2).
\]
This analytical correction modifies nonlogarithmic source
descendants but does not change the first-order
$r^{-2}\log(r_0/r)$ coefficient. Consequently, the
transformation of the leading source amplitudes and
quadrupolar logarithmic coefficients used above remains
applicable and yields
\begin{equation}
C^{(F^2)}_{\rm can}=a
\begin{pmatrix}0&6Q^3/5\\6Q^3/5&-27MQ^2/5\end{pmatrix}.
\label{eq:F2-canonical-matrix}
\end{equation}
This also agrees with the one-invariant result of Ref.~\cite{Barbosa:2026qcv} after translating between conventions. For reference, its action parameters obey $\kappa^2=8\pi$, $e^2=4\pi$, $c_2=a/(128\pi)$, $\kappa^2M_B=2M$, and $Q_B=-\sqrt2 Q/\kappa$ when $F_{rt}=eQ_B/r^2$.
Deriving both contributions in the same static variables
fixes their relative normalization and sign before
they are combined.

Combining Eqs.~\eqref{eq:F2-canonical-matrix} and~\eqref{eq:canonEH} yields
\begin{equation}
C^{\rm EH}_{\rm can}=\frac{9aQ^2}{10}
\begin{pmatrix}0&-Q\\-Q&M\end{pmatrix}+O(a^2).
\label{eq:EH-canonical-matrix}
\end{equation}
The matrix multiplies $r^{-2}\log(r_0/r)$ and has dimensions of length$^5$. Its dimensionless version is $C^{\rm EH}_{\rm can}/[r_+^{(0)}]^5$; using the corrected horizon changes this normalization only at higher order. The mixed coefficient is $6/5-21/10=-9/10$, and the electromagnetic diagonal coefficient is $-27/5+63/10=9/10$. Both cancellations are therefore required by the full quartic action. In particular,
\begin{equation}
C_{11}^{\rm EH}=0,\qquad
\frac{|C_{12}^{\rm EH}|}{|C_{22}^{\rm EH}|}=\frac{|Q|}{M}
\quad (aQ\ne0).
\label{eq:complete_ratio}
\end{equation}
The first identity concerns the logarithmic coefficient at this order; it does not determine a finite gravitational Love coefficient.

\subsection{Direct numerical verification of the logarithmic response}
\label{sec:numerical_check}
As an independent check of the asymptotic calculation, we integrate the
complete static perturbation equations without expanding their radial
coefficients in inverse powers of \(r\). To isolate the contribution
linear in the Euler-Heisenberg coupling, we introduce an auxiliary
parameter \(\lambda\) according to
\begin{equation}
 f_{\lambda}(r)
 =
 1-\frac{2M}{r}+\frac{Q^{2}}{r^{2}}
 -\lambda\frac{Q^{4}}{20r^{6}}\,,
 {\cal L}_{F}
 =
 1+\lambda\frac{Q^{2}}{2r^{4}}\,,
 {\cal L}_{GG}
 =
 -\frac{7\lambda}{4}\,.
 \label{eq:numerical_deformation}
\end{equation}
The physical Euler-Heisenberg correction is recovered by identifying
\(\lambda=a\) at linear order.
For each value of \(\lambda\), the outer horizon \(r_{+}(\lambda)\) is
recomputed from
\begin{equation}
 f_{\lambda}\bigl(r_{+}(\lambda)\bigr)=0.
 \label{eq:numerical_horizon}
\end{equation}
Two linearly independent horizon-regular solutions are then initialized
with a Frobenius expansion at
\(r=r_{+}(\lambda)+\epsilon\) and propagated to the asymptotic region.
The integrations are performed in the variable
\begin{equation}
 t=\log\bigl(r-r_{+}(\lambda)\bigr),
 \label{eq:numerical_coordinate}
\end{equation}
which resolves the near-horizon region while allowing the solutions to
be followed over a large radial interval.
At large radius, both perturbation fields and their first derivatives
are fitted simultaneously to the corresponding power-law and
power-logarithmic asymptotic bases. For the two independent regular
solutions, the coefficients of the growing tidal fields define a source
matrix \(S(\lambda)\), while the coefficients multiplying
\(r^{-2}\log(r_{0}/r)\) define a logarithmic-response matrix
\(B(\lambda)\). The response matrix in the raw numerical normalization is therefore
\begin{equation}
 C_{\mathrm{raw}}(\lambda)
 =
 B(\lambda)S^{-1}(\lambda).
 \label{eq:numerical_raw_matrix}
\end{equation}
After converting the gravitational source to the canonical convention
used in the analytic calculation, one obtains
\begin{equation}
 C_{\mathrm{can}}(\lambda)
 =
 C_{\mathrm{raw}}(\lambda)
 \begin{pmatrix}
 -4 & 0\\
 0 & 1
 \end{pmatrix}.
 \label{eq:numerical_canonical_matrix}
\end{equation}
The term linear in the Euler-Heisenberg coupling is extracted by the
centered finite difference
\begin{equation}
 C_{\mathrm{can}}^{(1)}
 =
 \left.
 \frac{\partial C_{\mathrm{can}}}{\partial\lambda}
 \right|_{\lambda=0}
 \simeq
 \frac{
 C_{\mathrm{can}}(\delta\lambda)
 -
 C_{\mathrm{can}}(-\delta\lambda)}
 {2\delta\lambda}.
 \label{eq:numerical_derivative}
\end{equation}
Table~\ref{tab:direct_numerical_check} compares the direct numerical
result with the analytic prediction for \(M=1\),
\(\delta\lambda=3\times10^{-4}\), \(\epsilon=10^{-6}\), and an
asymptotic fitting interval \(50\leq r\leq120\). The numerical values
reproduce
\begin{equation}
 C_{\mathrm{can}}^{(1)}
 =
 \frac{9Q^{2}}{10}
 \begin{pmatrix}
 0 & -Q\\
 -Q & 1
 \end{pmatrix},
 \label{eq:numerical_expected_matrix}
\end{equation}
with relative differences below \(2\times10^{-7}\) for every nonzero
matrix element.
\begin{table}[t]
\centering
\caption{
Direct numerical determination of the logarithmic-response matrix.
The displayed entries are the coefficients linear in \(\lambda\).
The analytic values are
\(C_{12}^{(1)}=C_{21}^{(1)}=-9Q^{3}/10\) and
\(C_{22}^{(1)}=9Q^{2}/10\).
}
\label{tab:direct_numerical_check}
\begin{ruledtabular}
\begin{tabular}{c c c c c}
\(Q\) &
\(C_{11}^{(1)}\) &
\(C_{12}^{(1)}\) &
\(C_{21}^{(1)}\) &
\(C_{22}^{(1)}\)
\\
\hline
\(0.2\) &
\(-9.48\times10^{-11}\) &
\(-0.0071999993\) &
\(-0.0071999989\) &
\(0.0359999944\)
\\
\(0.4\) &
\(-1.46\times10^{-9}\) &
\(-0.0575999948\) &
\(-0.0575999916\) &
\(0.1439999776\)
\\
\(0.6\) &
\(-6.92\times10^{-9}\) &
\(-0.1943999832\) &
\(-0.1943999742\) &
\(0.3239999504\)
\\
\(0.8\) &
\(-1.98\times10^{-8}\) &
\(-0.4607999629\) &
\(-0.4607999469\) &
\(0.5759999154\)
\end{tabular}
\end{ruledtabular}
\end{table}
The vanishing entry \(C_{11}^{(1)}=0\) is recovered up to a small
absolute numerical residual. In addition, the independently extracted
off-diagonal components satisfy reciprocity,
\(C_{12}^{(1)}=C_{21}^{(1)}\), to better than two parts in
\(10^{7}\). As a null test, repeating the asymptotic extraction at
\(\lambda=0\) gives logarithmic coefficients smaller than
\(6\times10^{-18}\).
This calculation provides a direct numerical verification of the full
Euler-Heisenberg logarithmic-response matrix, including the
contribution of the second electromagnetic invariant. It verifies the
coefficient of the local logarithmic running. It does not determine the
finite, scale-dependent tidal response, which additionally requires
global matching between the horizon-regular and asymptotic solutions.

\subsection{Analytic charge dependence}
\label{sec:numerical_G2}
We substitute the closed-form RN seeds into both field equations, check their horizon behavior, compute the first-order forcing for each invariant independently, and solve the resonant coefficients exactly. The result is rational in the symbolic parameters; no least-squares extraction or fitted charge exponent is required. In particular, the $Q^3$ mixed dependence and $MQ^2$ electromagnetic diagonal dependence follow algebraically.

\begin{figure}[t]
\centering
\includegraphics[width=0.45\textwidth]{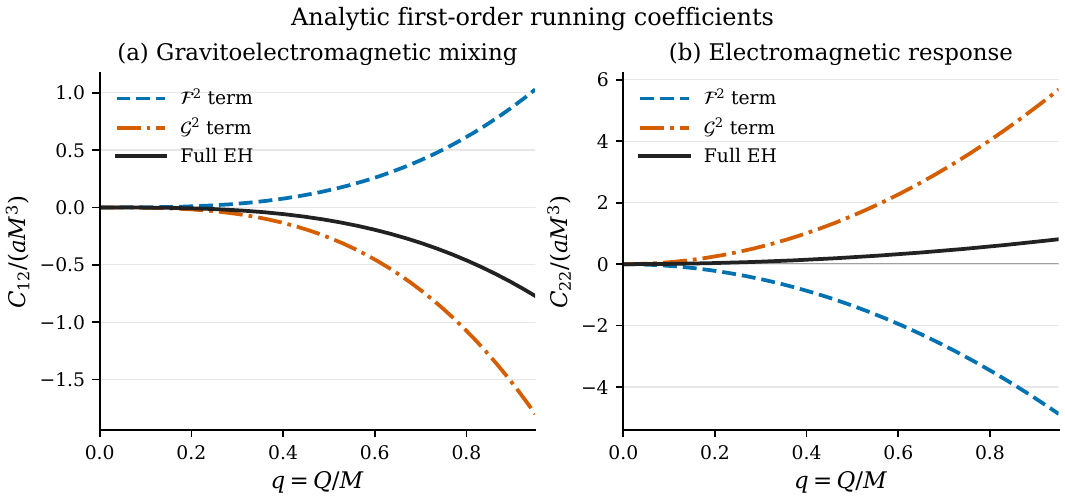}
\caption{Analytic quadrupolar logarithmic coefficients in the convention $\log(r_0/r)$, with $q=Q/M\ge0$ and normalization $C/(aM^3)$. Left: signed mixed entry; right: electromagnetic diagonal entry. The $F^2$ and $G^2$ contributions partially cancel. These curves are exact evaluations of the first-order formulas, not a numerical integration of the horizon boundary-value problem.}
\label{fig:running_components}
\end{figure}

Fig.~\ref{fig:running_components} shows these analytic coefficients. Separates the first invariant, the second invariant, and their sum, since displaying only the isolated $G^2$ contribution can obscure the cancellation in the complete Euler-Heisenberg theory.

Horizon-series data for a future global extraction in the
isolated second-invariant problem are collected in
Appendix~\ref{app:horizon-data}. They are not required
for the local logarithmic coefficients derived here.

\subsection{Canonical channels and interpretation
}
\label{subsec:physical_interpretation}
In the canonical basis, index 1 labels the axial gravitational source and index 2 the electromagnetic source. The symmetric mixed entries describe the reciprocal logarithmic response between these sources. The second invariant alone gives $|C_{12}|/|C_{22}|=|Q|/(3M)$, whereas their Euler-Heisenberg sum gives $|Q|/M$. The cancellation therefore changes the relative importance of mixing.

The vanishing of $C_{11}$ in both pieces and their sum is a statement about the leading logarithmic running. It neither sets the finite gravitational coefficient to zero nor removes the gravitational response to an electromagnetic source. The eigenvalues of the complete matrix are
\begin{equation}
\lambda_\pm=\frac{9aQ^2}{20}
\left(M\pm\sqrt{M^2+4Q^2}\right).
\label{eq:EH_eigenvalues}
\end{equation}
For $a>0$ and $Q\neq0$, these eigenvalues have opposite
signs. They belong to the logarithmic running matrix,
not to an energy Hessian or a mode-frequency operator,
and therefore do not imply instability. For fixed
bulk couplings, the subtraction-scale dependence resides
in the finite response matrix, whose running is given
by Eq.~\eqref{eq:renorm2}. Charge conjugation
$Q\to -Q$ reverses the mixed entry while leaving
the diagonal entries and eigenvalues unchanged.

\subsection{One logarithm and renormalization convention}
\label{sec:renormalization}
Throughout, $C$ is the coefficient of $\log(r_0/r)$. To use an inverse-length scale $\mu=1/r_0$, write the decaying part of a solution as
\begin{equation}
\bm\Psi_{\rm dec}=r^{-2}
\left[\mathbf B(\mu)-C\mathbf A\log(\mu r)\right]+\cdots.
\label{eq:renorm1}
\end{equation}
Keeping the physical solution fixed gives
\begin{equation}
\mathbf B(e^s\mu)=\mathbf B(\mu)+s C\mathbf A,
\qquad \frac{d\Lambda}{d\log\mu}=C.
\label{eq:renorm2}
\end{equation}
Here, $\Lambda$ is the finite matrix that relates $\mathbf B$ to $\mathbf A$ in the stated basis. If the logarithm displayed instead is $+\log(\mu r)$, its coefficient is $-C$. A finite counterterm shifts $\Lambda$; it does not change the leading logarithmic coefficient within the fixed operator and field basis. In particular, $\Lambda$ and the logarithm must be kept together in a matched observable. A convenient reference is $\mu=1/r_+^{(0)}$, but choosing that scale is not a calculation of the finite response.

\section{Distinguishing theories with the same electric geometry}
\label{sec:degeneracy}
\subsection{Independent quartic couplings}
Consider the two-parameter deformation
\begin{equation}
L(F,G)=F-\frac a2 F^2-\frac b2 G^2+O(a^2,ab,b^2),
\label{eq:ab_action}
\end{equation}
where Euler-Heisenberg corresponds to $b=7a/4$. At fixed $a,M,Q$, varying $b$ changes neither the electric first integral nor the background lapse: $G=L_G=0$ in that solution. However, the static axial operator depends on $L_{GG}=-b$. Combining the independently verified recursions gives
\begin{equation}
C(a,b)=\frac{Q^2}{5}
\begin{pmatrix}
0&6(a-b)Q\\
6(a-b)Q&(-27a+18b)M
\end{pmatrix}.
\label{eq:general_matrix}
\end{equation}
This equation makes explicit the degeneracy of the background. Within this specified two-operator model, an electric metric fixes only the first-invariant deformation, while the axial running also depends on the second invariant. A particularly transparent example is obtained by setting
$a=0$ while keeping $b\neq0$. The electric background
then coincides with Reissner-Nordstr\"om, whereas
Eq.~\eqref{eq:general_matrix} gives
\begin{equation}
C(0,b)
=
\frac{6bQ^2}{5}
\begin{pmatrix}
0 & -Q\\
-Q & 3M
\end{pmatrix}
+\mathcal O(b^2).
\label{eq:RN_background_nonzero_running}
\end{equation}
Thus, within the two-operator theory displayed,
an undeformed Reissner-Nordstr\"om geometry does not
imply the Einstein-Maxwell logarithmic response. This example isolates the constitutive information
carried by the axial fluctuations.

For $Q\ne0$ define
\begin{equation}
X=\frac{5C_{12}}{6Q^3}=a-b,\qquad
Y=\frac{5C_{22}}{9MQ^2}=-3a+2b.
\end{equation}
The coefficient map has a determinant $-1$ and can be inverted:
\begin{equation}
a=-2X-Y,\qquad b=-3X-Y.
\label{eq:coupling_inverse}
\end{equation}
These are algebraic identifiability relations for known $M,Q$ and canonically normalized logarithmic coefficients. They are not an observational error forecast. Although the normalized coefficient map remains nonsingular, the unnormalized mixed and diagonal responses vanish as $Q^3$ and $Q^2$, respectively. Consequently, sensitivity to the two couplings is progressively lost toward the neutral limit. Extra EFT operators would, in general, enlarge the parameter space and spoil this two-parameter inversion.

The cancellation lines are different. At $b=a$, the mixed running vanishes, but at $C_{22}=-9aMQ^2/5$. At $b=3a/2$, the electromagnetic diagonal vanishes, but at $C_{12}=-3aQ^3/5$. For a nonzero charge, both nonzero channels vanish simultaneously only if $a=b=0$. The Euler-Heisenberg point obeys
\begin{equation}
X=-\frac{3a}{4},\qquad Y=\frac a2,\qquad
\frac{C_{12}^{\rm EH}}{Q^3}+\frac{C_{22}^{\rm EH}}{MQ^2}=0.
\label{eq:EH_null_relation}
\end{equation}
This model-specific relation is a useful check for any direct numerical extraction performed under the same conventions.

\begin{figure}[H]
\centering
\includegraphics[width=0.45\textwidth]{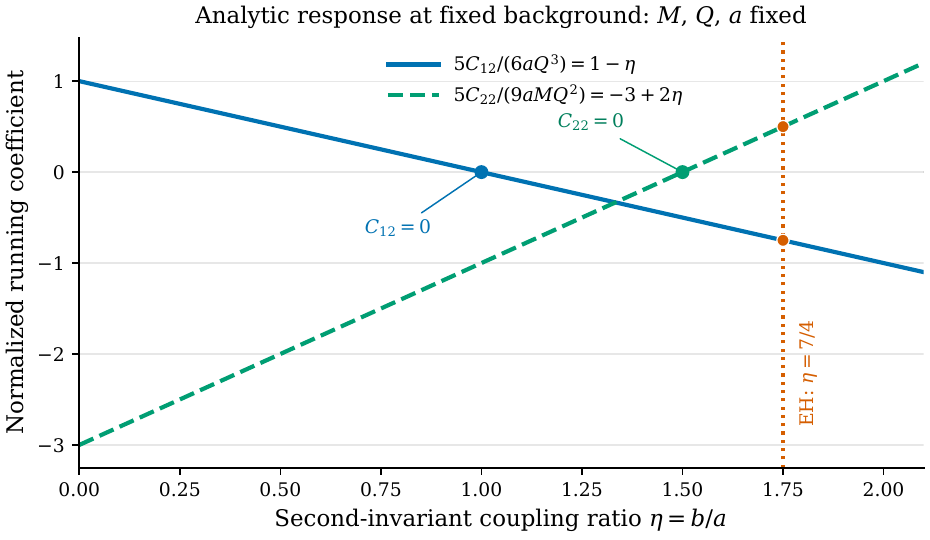}
\caption{Analytic response for electric backgrounds with identical $a,M,Q$ and different $\eta=b/a$, for $a\ne0$ and $Q\ne0$. The curves show $X/a=1-\eta$ and $Y/a=-3+2\eta$. Their zeros occur at different coupling ratios; the Euler-Heisenberg value $\eta=7/4$ yields $X/a=-3/4$ and $Y/a=1/2$.}
\label{fig:coupling_diagnostic}
\end{figure}

The exterior perturbative control is
\begin{equation}
\frac{\max(|a|,|b|)Q^2}{[r_+^{(0)}]^4}\ll1.
\end{equation}
Fig.\ref{fig:coupling_diagnostic} illustrates these relations as functions of $\eta=\frac{b}{a}$, highlighting the distinct cancellation points of the two running coefficients and the Euler–Heisenberg value $\eta=\frac{7}{4}$.

For example, the axial electromagnetic angular factor and its ratio to the kinetic factor are
\begin{align}
K_\Omega&=L_F+E^2L_{GG}
=1+\left(\frac a2-b\right)\frac{Q^2}{r^4}+\cdots,\\
\frac{K_\Omega}{L_F}&=1-b\frac{Q^2}{r^4}+\cdots.
\label{eq:angular_factor}
\end{align}
Their positivity holds in the perturbative domain. This is a local check on the displayed axial operator, not a proof of mode stability of the coupled black-hole system. The sufficient simple-root control condition ~\eqref{eq:horizon_control} also remains in force.

\subsection{Comparison with the extremal two-invariant result}
\label{subsec:extremal_literature_check}

The second invariant already contributes to the extremal response
calculated by Noumi and Wong~\cite{Noumi:2026shc}.  Their action contains
$\alpha_2\kappa^4(F^{\rm NW}_{\mu\nu}{}^\star
F_{\rm NW}^{\mu\nu})^2$, with Maxwell normalization $-F_{\rm NW}^2/4$.
In our units, $\kappa^2=8\pi$ and
$F^{\rm NW}_{\mu\nu}=F_{\mu\nu}/\sqrt{4\pi}$.  Comparing this term
with $-\gamma G^2/(8\pi)$, where $\gamma=L_{GG}$, gives
\begin{equation}
 \gamma=-64\kappa^2\alpha_2.
 \label{eq:NW_gamma_map}
\end{equation}
Thus, the Euler-Heisenberg value $\gamma=-7a/4$ corresponds to
$\alpha_2=7a/(256\kappa^2)$.

Set $Q=M=r_h>0$ and write $x=r/r_h$. The dimensionless matrix multiplying $x^{-2}\log x$ differs from our matrix multiplying $r^{-2}\log(r_0/r)$ by an overall minus sign and a factor $r_h^{-5}$. There is also a relative gravitational-field sign between the two conventions.  Indeed, their canonical pair is $(\mathcal Q/\kappa,a_{\rm NW})$, and their pure gravitational source induces an electromagnetic seed proportional to $-(x^2-1)$; our convention has the opposite relative sign. With
$D=\mathrm{diag}(-1,1)$, the source-response transformation is
\begin{align}
 \widehat C_{\rm NW}^{(\alpha_2)}
 &=-\frac{1}{r_h^5}
 D\left.\Delta C_{\rm can}^{(G^2)}\right|_{Q=M=r_h}D
 \nonumber\\
 &=\frac{\kappa^2\alpha_2}{r_h^2}
 \begin{pmatrix}
 0&-384/5\\[-1mm]
 -384/5&-1152/5
 \end{pmatrix}.
 \label{eq:NW_exact_comparison}
\end{align}
This agrees entry by entry with the terms proportional to $\alpha_2$
in Eqs.~(86) and~(88) of Ref.~\cite{Noumi:2026shc}, after setting
$\alpha_1=\alpha_3=0$.  More generally, their full quartic result with
$\alpha_3=0$ is
\begin{equation}
 \widehat C_{\rm NW}^{(\alpha_1,\alpha_2)}
 =\frac{\kappa^2}{5r_h^2}
 \begin{pmatrix}
 0&384(\alpha_1-\alpha_2)\\
 384(\alpha_1-\alpha_2)&1728\alpha_1-1152\alpha_2
 \end{pmatrix}.
 \label{eq:NW_full_quartic_comparison}
\end{equation}
The first-invariant normalization is $a=64\kappa^2\alpha_1$. Consequently, the Euler-Heisenberg ratio $\alpha_2=7\alpha_1/4$ also reproduces the extremal value of our complete two-invariant matrix after the same change of field and logarithm conventions.

The agreement provides a check of the relative field
normalization, the coupling dictionary, and the logarithm
convention. The present expressions additionally determine
the dependence on independent subextremal $M$ and $Q$.
This comparison concerns the local logarithmic coefficients;
it does not establish a uniform extremal limit of the
finite horizon-matched response.

\section{Conclusions}
\label{sec:conclusions}

The two electromagnetic invariants of the minimally coupled quartic Euler-Heisenberg action affect axial tidal response differently. The first deforms the electric background and perturbation operator; the second leaves the background unchanged but enters the quadratic axial action. Using the coupled static equations and a common canonical normalization, we obtained both quadrupolar logarithmic contributions directly from their resonant asymptotic recursion. Their sum is Eq.~\eqref{eq:EH-canonical-matrix}, with $C_{11}=0$, reciprocal mixed entries proportional to $aQ^3$, and an electromagnetic diagonal proportional to $aMQ^2$.

The complete Euler-Heisenberg result exhibits a partial cancellation between the two quartic contributions in both nonzero channels. In the fixed canonical basis and logarithm convention adopted here,
including the second invariant reverses the signs of the mixed and
electromagnetic diagonal coefficients relative to the first-invariant
contribution alone. The absolute mixed-to-diagonal ratio becomes $|Q|/M$, showing that mixing grows in relative importance as the magnitude of the charge increases at fixed mass. Therefore, neglecting an interaction that is invisible to the electric background changes the logarithmic response at the same perturbative order as the background deformation.

The exact horizon-regular Reissner-Nordstr\"om seeds and the recovered Schwarzschild solution provide nontrivial checks of the static equations and the identification of source and response. In particular, a decaying radial power in the reconstructed metric
perturbation can belong to the growing source branch and need not
represent an independent Love coefficient. The quadratic action also
fixes the relative normalization of the gravitational and
electromagnetic channels, so the symmetry of the canonical logarithmic
matrix follows from the coupled theory rather than from an imposed
relation between unnormalized amplitudes.

The local logarithmic coefficients retain the full
dependence on subextremal $Q/M$ at first order in the
displayed couplings. Their extremal local limit agrees
with the corresponding two-invariant EFT result after
convention conversion, providing a consistency check
on the generic-charge expressions. The independent-coupling
matrix in Eq.~\eqref{eq:general_matrix} and the inverse
relations in Eq.~\eqref{eq:coupling_inverse} then make
explicit how the axial running resolves a degeneracy
of the electric background within this restricted model.

Within this restricted two-operator model, the mixed and
electromagnetic diagonal entries probe linearly independent
combinations of the couplings. Their cancellation conditions are
different, and at nonzero charge both coefficients vanish
simultaneously only in the Maxwell limit. The inverse relations
therefore establish algebraic identifiability for specified $M$,
$Q$, and canonical normalization; they do not constitute an
observational reconstruction. This distinction becomes particularly
relevant near the neutral limit, where the unnormalized responses vanish and sensitivity to the two couplings is progressively lost. Additional effective
operators would enlarge the parameter space and could invalidate
this two-parameter inversion.

The quantity determined in this work is the coefficient of logarithmic
running, not the finite horizon-matched Love matrix. A change of
subtraction scale is compensated by a shift of the finite coefficients,
leaving the matched solution unchanged; both contributions must
therefore be retained in a physical response. Consequently,
$C_{11}=0$ excludes only the leading gravitational diagonal running
at this order: it neither fixes the finite gravitational Love
coefficient nor eliminates the gravitational response to an
electromagnetic source. Likewise, the finite extremal limit of the
local coefficients does not establish a uniform extremal limit of
the horizon-matched problem. The perturbative horizon construction
remains subject to the weak-field and simple-root conditions
specified in the analysis.

A complete QED prediction would additionally require curvature and
derivative operators and a physical field-strength hierarchy.
These distinctions separate the results proved here from the
calculations needed for finite response and phenomenological
applications.
 A natural next step is to determine the finite response by propagating
two independent horizon-regular solutions and matching them to an
asymptotic basis that retains both the logarithmic terms and all
relevant source descendants. The closed-form running coefficients
derived here provide analytic benchmarks for that extraction,
including its normalization and scale dependence. Within the present
electrovacuum setting, our results thus show concretely how static
axial tidal running retains information about the electromagnetic
action that the equilibrium geometry alone cannot supply.

\begin{acknowledgments}
Y. V. acknowledges support by the Direcci\'on de Investigaci\'on y Desarrollo de la Universidad de La Serena, Grant No. PR25538511.
\end{acknowledgments}

\appendix

\section{Derivation of the static axial equations and canonical normalization}
\label{app:axial_derivation}

This appendix collects the ingredients needed to verify the static
odd-parity system used in the main text and to make explicit the origin
of the $ L_{ GG}$ term.  The purpose is not to repeat
the full time-dependent nonlinear-electrodynamics perturbation
formalism, but to display the projections and field redefinitions that
are directly relevant for Eqs.~\eqref{eq:static-gravitational}-\eqref{eq:static-electromagnetic} and for
the canonical matrix in Eq.~\eqref{eq:canonEH}.

\subsection{Linearized Einstein-NLED equations}

For the conventions of Eqs.~\eqref{eq:FG_def}-\eqref{eq:action_EEH},
the electromagnetic field equation can be written as
\begin{equation}
\nabla_\mu\!\left(
 L_FF^{\mu\nu}
+L_{G}\,{}^\star F^{\mu\nu}
\right)=0.
\label{app:maxwell}
\end{equation}
The stress tensor may equivalently be expressed as
\begin{equation}
T_{\mu\nu}
=\frac{1}{4\pi}
\left[
 L_{F}F_{\mu\rho}F_\nu{}^\rho
+g_{\mu\nu}
\left(
L_{ G}G- L
\right)
\right],
\label{app:stress}
\end{equation}
where the four-dimensional identity between
$F_{\mu\rho}{}^\star F_\nu{}^\rho$ and $G g_{\mu\nu}$ has
been used.  The background is purely electric,
$\bar F_{tr}=E(r)$, with
\begin{equation}
\bar{ G}=0,
\qquad
\bar{L}_{ G}=0,
\label{app:Gbackground}
\end{equation}
for the parity-even Euler-Heisenberg Lagrangian.

In Regge-Wheeler gauge the static odd perturbations are
\begin{equation}
\delta g_{tA}=h_0 S_A,
\qquad
\delta g_{rA}=h_1 S_A,
\qquad
\delta A_A=u S_A,
\label{app:oddansatz}
\end{equation}
where the $(\ell,m)$ labels are suppressed.  We use the standard
axial-harmonic identities
\begin{align}
D^A S_A&=0,
\qquad L=\ell(\ell+1),
\nonumber\\
D_A S_B-D_B S_A&=L\,\epsilon_{AB}Y_{\ell m}.
\label{app:harmonicidentities}
\end{align}
The electromagnetic perturbation therefore satisfies
\begin{equation}
\delta F_{rA}=u'S_A,
\qquad
\delta F_{AB}=L\,u\,\epsilon_{AB}Y_{\ell m}.
\label{app:dFodd}
\end{equation}

Parity immediately simplifies the constitutive variations.  In the
odd sector one has $\delta F=0$, whereas
$\delta G$ is nonzero.  Consequently, on the electric
background,
\begin{equation}
\delta L_{ G}
=
 L_{ GG}\,\delta G.
\label{app:dLG}
\end{equation}
For the part proportional to $L_{GG}$ it is
sufficient, consistently through $\mathcal O(a)$, to evaluate
$\delta G$ on the Reissner-Nordstr\"om electric background:
any Euler-Heisenberg correction to the background electric field
would multiply $ L_{GG}=\mathcal O(a)$ and hence
enter only at $\mathcal O(a^2)$.  With the orientation convention
implicit in Eq.~\eqref{app:harmonicidentities},
\begin{equation}
\delta G
=
\frac{LQ}{r^4}\,u\,Y_{\ell m}
+\mathcal O(a).
\label{app:dGexplicit}
\end{equation}
Thus the apparently inactive pseudoscalar invariant generates a
linear constitutive perturbation even though
$\bar{G}=0$.

The static angular equation gives $(fh_1)'=0$. More strongly, the $rA$ Einstein equation reduces to $(L-2)h_1=0$ after the background equations are used. Thus $h_1=0$ for $\ell\ge2$. The stationary dipole requires a separate treatment and is not part of the tidal calculation here.

The remaining independent projections are the $tA$ Einstein equation
and the axial projection of Eq.~\eqref{app:maxwell}.  The background
curvature terms in the $tA$ projection must be retained.  Introducing
$m(r)=r[1-f(r)]/2$, the gravitational projection is
\begin{equation}
h_0''
-\frac{1}{f}
\left[
\frac{L}{r^2}
-\frac{4m(r)}{r^3}
\right]h_0
=
\frac{4Q}{r^2}u'
+O(a^2).
\label{app:h0projected}
\end{equation}

Using Eq.~\eqref{eq:EEH_f},
\begin{equation}
\frac{L}{r^2}-\frac{4m(r)}{r^3}
=
\frac{L}{r^2}-\frac{4M}{r^3}+\frac{2Q^2}{r^4}
-\frac{aQ^4}{10r^8}+\mathcal O(a^2),
\label{app:gravpotentialexpanded}
\end{equation}
which is Eq.~\eqref{eq:static-gravitational}.  In particular, for $Q=a=0$ and
$L=6$ the solution $h_0=r^3-2Mr^2$ makes the left-hand side vanish
identically.  This check is sensitive to the $-4M/r^3$ curvature term
and is the simplest way of detecting its omission.

The axial Maxwell projection gives
\begin{equation}
\left(L_F f u'\right)'
-\frac{L L_F}{r^2}u
-\frac{LQ^2L_{GG}}{r^6}u
=
\frac{Q}{r^2}
\left(
h_0'-\frac{2h_0}{r}
\right).
\label{app:uprojected}
\end{equation}
In deriving Eqs.~\eqref{app:h0projected} and~\eqref{app:uprojected}, we used the exact
background first integral
\begin{equation}
r^2 L_F E=Q.
\end{equation}
Consequently, the factors $L_F E$ appearing in the
gravitoelectromagnetic mixing terms reduce to $Q/r^2$.
No additional factor of $L_F$ acts on either $u'$ or $h_0$
in these terms. Equations~\eqref{app:h0projected} and~\eqref{app:uprojected} are
precisely Eqs.~\eqref{eq:static-gravitational} and~\eqref{eq:static-electromagnetic}.

The origin of the last term on the first line of
Eq.~\eqref{app:uprojected} is especially simple.  Combining
Eqs.~\eqref{app:dLG} and~\eqref{app:dGexplicit}, the perturbation
$\delta L_{ G}$ multiplies the dual of the background
electric field in Eq.~\eqref{app:maxwell}.  The angular derivative
then converts the scalar harmonic back into $S_A$ and produces
\begin{equation}
-\frac{LQ^2L_{ GG}}{r^6}\,u.
\label{app:GGterm}
\end{equation}
This is the term that would be lost if one inferred the perturbation
problem from the background metric alone or truncated the theory to
$ L( F)$.

Before expansion, the angular term is $-L E^2L_{GG}u/r^2=-LQ^2L_{GG}u/(r^6L_F^2)$. Replacing $L_F^{-2}$ by one in the term proportional to $L_{GG}$ is valid only to first order in the quartic couplings. The static equations in the main text are understood with this consistent truncation.

For Euler-Heisenberg,
\begin{equation}
L_{F}
=
1+\frac{aQ^2}{2r^4}+\mathcal O(a^2),
\qquad
 L_{ GG}
=
-\frac{7a}{4},
\label{app:EHderivatives}
\end{equation}
so the angular electromagnetic operator contains
\begin{align}
r^4 L_{F}
+Q^2 L_{GG}
&=
r^4+\frac{aQ^2}{2}-\frac{7aQ^2}{4}
+\mathcal O(a^2)
\nonumber\\
&=
r^4-\frac{5aQ^2}{4}
+\mathcal O(a^2),
\label{app:EHcombination}
\end{align}
which reproduces Eq.~\eqref{eq:EHoperator}.  The
$G^2$ contribution is therefore of the same perturbative
order as the $ F^2$ contribution and cannot be discarded in a
consistent Euler-Heisenberg axial calculation.

\subsection{Canonical normalization and reciprocity}

The relative normalization of the two channels is fixed by the
quadratic Einstein-NLED action. We use real scalar harmonics with
$\int Y_{\ell m}^{2}d\Omega=1$, so that
$\int S_A S^A d\Omega=L$, and define
\begin{equation}
\lambda=L-2=(\ell-1)(\ell+2),\qquad
\alpha=\sqrt{\frac{4\pi}{L}},\qquad \ell\geq2.
\label{app:canonical-conventions}
\end{equation}
With these conventions, the electromagnetic kinetic term before
canonical normalization has coefficient $L/(8\pi)$.
 With this choice of $\alpha$, the kinetic term has coefficient $1/2$. Setting
$\alpha=1$ instead amounts to retaining a common prefactor in the
quadratic action and has no effect on the response matrix.

To define the master variables locally, restore the time dependence
and introduce
\begin{equation}
\chi=\dot h_1-h_0'+\frac{2h_0}{r},\qquad
s(r)=\sqrt{L_F(r)}.
\label{app:local-chi}
\end{equation}
For the parity-even electric NLED theories considered here,
$L_G=L_{FG}=0$ at $G=0$. Assuming $L_F>0$, the action-normalized
variables are
\begin{equation}
\Psi_h=\frac{r\chi+4Qu/r}{2\alpha\sqrt\lambda},
\qquad
\Psi_a=\frac{s u}{\alpha}.
\label{app:local-canonical-fields}
\end{equation}
The charge in this expression is the conserved charge
$Q=r^2L_FE$. The electromagnetic sign follows from our convention
$F_{tr}=+E$. In particular, it must be translated when comparing
with conventions that use $F_{tr}=-E$.

The Einstein constraints give the reconstruction
\begin{align}
h_1&=-\frac{2\alpha r}{f\sqrt\lambda}\dot\Psi_h,
\nonumber\\
h_0&=-\frac{2\alpha f}{\sqrt\lambda}(r\Psi_h)',
\qquad
u=\frac{\alpha}{s}\Psi_a.
\label{app:general-reconstruction}
\end{align}
The local definition in Eq.~\eqref{app:local-canonical-fields}
is essential: the equation for $h_0$ alone would leave an undetermined
addition $C/r$ to $\Psi_h$. Such an addition is not an independent
freedom once the definition and the field equations are imposed.
Under an odd diffeomorphism, the variations of $\dot h_1$,
$h_0'$, and $2h_0/r$ cancel in $\chi$. The transverse electromagnetic
amplitude $u$ is invariant under both odd diffeomorphisms of the
electric background and electromagnetic gauge transformations.
Thus the master fields are gauge invariant before taking the
static limit. At zero frequency, the $rA$ Einstein equation also
gives $\lambda h_1=0$, consistently with the reconstruction.

After eliminating the nondynamical variables and discarding
boundary terms, the bulk quadratic action becomes
\begin{equation}
\begin{split}
S_{\rm odd}^{(2)}=\frac12\sum_{\ell m}\int dt\,dr_*
\big[&\dot{\boldsymbol\Psi}^{\,T}\dot{\boldsymbol\Psi}
-(\partial_{r_*}\boldsymbol\Psi)^T
 (\partial_{r_*}\boldsymbol\Psi)\\
&-\boldsymbol\Psi^T\mathbf V_\ell\boldsymbol\Psi\big],
\qquad
\boldsymbol\Psi=\begin{pmatrix}\Psi_h\\\Psi_a\end{pmatrix},
\end{split}
\label{app:canonical-action}
\end{equation}
where $dr_*/dr=f^{-1}$. For a general parity-even electric NLED
background satisfying the Einstein equations, the potential entries
are
\begin{align}
V_{hh}&=f\left(\frac{\lambda+2f}{r^2}-\frac{f'}r\right),
\nonumber\\
V_{ha}=V_{ah}&=-\frac{2Q\sqrt\lambda\,f}{r^3\sqrt{L_F}},
\nonumber\\
V_{aa}&=f\left[
\frac{L}{r^2}\left(1+\frac{E^2L_{GG}}{L_F}\right)
+\frac{4Q^2}{r^4L_F}\right]
+\frac{\partial_{r_*}^{2}\sqrt{L_F}}{\sqrt{L_F}}.
\label{app:general-canonical-potential}
\end{align}
These expressions exhibit the symmetric gravitoelectromagnetic
coupling explicitly and agree with the general NLED perturbation
structure after translating electric-field and charge conventions
\cite{Daghigh:2021psm,Barbosa:2026qcv}.
For the Euler-Heisenberg truncation, every product in
Eq.~\eqref{app:general-canonical-potential} is expanded through
$\mathcal O(a)$. In particular,
$s=1+aQ^2/(4r^4)+\mathcal O(a^2)$.
This analytic field normalization changes nonlogarithmic source
descendants but cannot change the first-order $r^{-2}\log r$
coefficient in the quadrupolar channel.

For the isolated second-invariant contribution, set $L_F=1$,
$L_{GG}=\gamma$, and keep
$f=1-2M/r+Q^2/r^2$. Equation~\eqref{app:general-canonical-potential}
then reduces to
\begin{eqnarray}
&& V_{hh}=f\left(\frac{L}{r^2}-\frac{6M}{r^3}+\frac{4Q^2}{r^4}\right),\\
&& V_{ha}=V_{ah}=-\frac{2Q\sqrt\lambda f}{r^3},\\
&& V_{aa}=f\left(\frac{L}{r^2}+\frac{4Q^2}{r^4}+\frac{\gamma LQ^2}{r^6}\right)
+O(\gamma^2).
\label{app:RN-gamma-potential}
\end{eqnarray}

The additional interaction acts directly in the electromagnetic
entry; the symmetric off-diagonal potential transmits its effect to
the gravitational channel.
For $\ell=2$, the static reconstruction becomes
\begin{equation}
h_0=-\alpha f(r\Psi_h)',\qquad
u=\alpha\Psi_a,\qquad h_1=0.
\label{eq:canonical-map}
\end{equation}

A useful check is the exact horizon-regular quadrupolar RN seed in
this basis:
\begin{equation}
\begin{aligned}
\Psi_h^{(0)}&=(r^3-Q^4/r)A_h+(Qr^2-Q^3)A_a,\\
\Psi_a^{(0)}&=(Qr^2-Q^3)A_h\\
&\quad+\left(r^3-\frac32Mr^2+\frac32MQ^2-\frac{Q^4}{r}\right)A_a.
\end{aligned}
\label{app:canonical-RN-seed}
\end{equation}

Direct substitution into the static equations obtained from
Eq.~\eqref{app:RN-gamma-potential} at $\gamma=0$ gives zero.
The seed has no $r^{-2}$ term. Consequently, the $r^{-2}$ term in the
raw metric solution does not represent an independent Maxwell Love
response: it is generated by reconstructing $h_0$ from the growing
canonical branch.

We now transform the isolated logarithmic response. Write
\begin{equation}
\Psi_I=A_Ir^3+\cdots+\mathcal K_Ir^{-2}\log(r_0/r)+\cdots.
\label{app:masterasymptotic}
\end{equation}

Here, $r_0$ is a reference length. Because $f=1+\mathcal O(r^{-1})$,
Eq.~\eqref{eq:canonical-map} implies
\begin{align}
-\alpha f(r\Psi_h)'&=-4\alpha A_h r^3
+\alpha \mathcal K_h r^{-2}\log(r_0/r)+\cdots,
\label{app:hmapexpanded}\\
\alpha\Psi_a&=\alpha A_a r^3
+\alpha \mathcal K_a r^{-2}\log(r_0/r)+\cdots.
\label{app:umapexpanded}
\end{align}
Thus the raw source amplitudes $(H,U)$ and logarithmic coefficients
$(b_h,b_u)$ satisfy
\begin{equation}
\begin{pmatrix}H\\U\end{pmatrix}
=\alpha\begin{pmatrix}-4&0\\0&1\end{pmatrix}
\begin{pmatrix}A_h\\A_a\end{pmatrix},\qquad
\begin{pmatrix}b_h\\b_u\end{pmatrix}
=\alpha\begin{pmatrix}\mathcal K_h\\\mathcal K_a\end{pmatrix}.
\label{app:source_response_map}
\end{equation}
The factor $-4$ acts on the growing gravitational source branch,
whereas the coefficient of the leading decaying logarithm has the
factor $+1$.

The raw recursion gives
\begin{equation}
\begin{aligned}
\begin{pmatrix}b_h\\b_u\end{pmatrix}
&=\gamma C_{\rm raw}\begin{pmatrix}H\\U\end{pmatrix},\\
C_{\rm raw}&=\begin{pmatrix}0&6Q^3/5\\-3Q^3/10&-18MQ^2/5\end{pmatrix}.
\end{aligned}
\label{app:Craw}
\end{equation}

Substituting Eq.~\eqref{app:source_response_map} cancels the common
factor $\alpha$ and yields
\begin{equation}
\begin{split}
\begin{pmatrix}\mathcal K_h\\\mathcal K_a\end{pmatrix}
&=\gamma C_{\rm raw}
\begin{pmatrix}-4&0\\0&1\end{pmatrix}
\begin{pmatrix}A_h\\A_a\end{pmatrix}\\
&=\gamma\begin{pmatrix}
0&6Q^3/5\\6Q^3/5&-18MQ^2/5
\end{pmatrix}
\begin{pmatrix}A_h\\A_a\end{pmatrix}.
\end{split}
\label{app:Ccanonical}
\end{equation}
The symmetric matrix therefore follows from the action-normalized
fields, rather than from an imposed equality of raw coefficients.
For $\gamma=-7a/4$, the second-invariant contribution is
\begin{equation}
\Delta C_{\rm can}^{(G^2)}
=a\begin{pmatrix}
0&-21Q^3/10\\-21Q^3/10&63MQ^2/10
\end{pmatrix}.
\label{app:CEHfinal}
\end{equation}
Equations~\eqref{app:Ccanonical} and~\eqref{app:CEHfinal} use
$\log(r_0/r)$; the coefficient changes sign when the logarithm is
written as $\log(r/r_0)$. The quadrupolar matrix has dimension
${\rm length}^5$, since $[a]=[\gamma]={\rm length}^2$ in the
geometric units used here. Its identification with dimensionless
Love coefficients additionally requires a declared length scale and
worldline matching normalization.

\subsection{Horizon data for the isolated second-invariant problem}
\label{app:horizon-data}

For a future global extraction, the isolated-$\gamma$ RN horizon data at $x=r-r_+^{(0)}$ may be organized as $h_0=a_1x+a_2x^2+\cdots$ and $u=b_0+b_1x+\cdots$. With $f_1=f_0'(r_+^{(0)})$ and $V_+=V_0(r_+^{(0)})$, the leading coefficients are
\begin{align}
b_1&=\frac{6([r_+^{(0)}]^4+Q^2\gamma)b_0
+Q[r_+^{(0)}]^4 a_1}{[r_+^{(0)}]^6 f_1},\\
a_2&=\frac12\left(\frac{V_+a_1}{f_1}
+\frac{4Qb_1}{[r_+^{(0)}]^2}\right).
\label{eq:horizon_data_checked}
\end{align}
Two independent regular seeds can then be propagated and matched to a source basis including its subleading terms. A numerical finite response requires variation of the start radius, horizon-series order, integration tolerance, outer matching radius, and asymptotic truncation.

\section{Asymptotic recursion and the resonant logarithm}
\label{app:asymptotic-recursion}

In this appendix we make explicit the asymptotic recursion leading to
the logarithmic matrix in Eq.~\eqref{eq:rawlog}. We isolate the second-invariant
deformation by keeping the Reissner-Nordstr\"om background fixed,
setting $L_F=1$, and treating
\[
\gamma\equiv L_{GG}
\]
as a perturbative parameter. We write
\begin{equation}
h_0=h_0^{(0)}+\gamma h_0^{(1)}+O(\gamma^2),
\qquad
u=u^{(0)}+\gamma u^{(1)}+O(\gamma^2).
\end{equation}
For $\ell=2$, $L=6$, the zeroth-order fields are the exact RN
solutions given in Eqs.~\eqref{eq:h0rec}-\eqref{eq:urec}. At first order in $\gamma$ one
obtains
\begin{align}
h_0^{(1)\prime\prime}
-\frac{1}{f}
\left(
\frac{6}{r^2}
-\frac{4M}{r^3}
+\frac{2Q^2}{r^4}
\right)h_0^{(1)}
&=
\frac{4Q}{r^2}u^{(1)\prime},
\label{eq:Bgrav}
\\
(fu^{(1)\prime})'
-\frac{6}{r^2}u^{(1)}
-\frac{Q}{r^2}
\left(
h_0^{(1)\prime}
-\frac{2}{r}h_0^{(1)}
\right)
&=
\frac{6Q^2}{r^6}u^{(0)} .
\label{eq:Bem}
\end{align}

At large radius the leading diagonal operator is
\begin{equation}
{\cal D}_0[y]=y''-\frac{6}{r^2}y ,
\end{equation}
whose indicial equation,
\begin{equation}
s(s-1)-6=0,
\end{equation}
has roots
\begin{equation}
s_+=3,\qquad s_-=-2.
\end{equation}
Thus the growing and decaying quadrupolar branches are respectively
$r^3$ and $r^{-2}$ and differ by five powers, $s_+-s_-=5=2\ell+1$.
The latter is the resonant response branch.

From Eq.~\eqref{eq:urec},
\begin{equation}
u^{(0)}
=
Ur^3-\frac14(QH+6MU)r^2
+\frac{Q^2}{4}(QH+6MU)
-\frac{UQ^4}{r},
\end{equation}
so that the inhomogeneous term in Eq.~\eqref{eq:Bem} is
\begin{equation}
\frac{6Q^2}{r^6}u^{(0)}
=
6Q^2U\,r^{-3}
-\frac32Q^2(QH+6MU)r^{-4}
+O(r^{-6}).
\label{eq:Bsource}
\end{equation}

We therefore use the asymptotic ansatz
\begin{align}
u^{(1)}
&=
b_{-1}r^{-1}
+b_u^{\log}r^{-2}\log\left(\frac{r_0}{r}\right)
+\cdots ,
\\
h_0^{(1)}
&=
b_h^{\log}r^{-2}\log\left(\frac{r_0}{r}\right)
+\cdots .
\end{align}
The finite $r^{-2}$ pieces are left unspecified because they belong to
the nonlogarithmic response and depend on the global matching
prescription.

For the electromagnetic diagonal operator,
\begin{equation}
{\cal D}_u[y]\equiv(fy')'-\frac{6}{r^2}y,
\end{equation}
the required identities are
\begin{align}
{\cal D}_u[r^{-1}]
&=
-4r^{-3}-6Mr^{-4}+O(r^{-5}),
\\
{\cal D}_u\left[
r^{-2}\log\left(\frac{r_0}{r}\right)
\right]
&=
5r^{-4}+O(r^{-5}\log r).
\end{align}
At order $r^{-3}$, Eq.~\eqref{eq:Bem} gives
\begin{equation}
-4b_{-1}=6Q^2U,
\end{equation}
hence
\begin{equation}
b_{-1}=-\frac32Q^2U.
\label{eq:bminusone}
\end{equation}

At order $r^{-4}$ the coefficient multiplying a pure $r^{-2}$ mode
vanishes because $s_-=-2$ is an indicial root. The recursion is
therefore resonant and the logarithmic term is required. At this order,
the contribution involving the gravitational correction on the right-hand side of
Eq.~\eqref{eq:Bem} starts only at $O(r^{-5}\log r)$, and one obtains
\begin{equation}
-6Mb_{-1}+5b_u^{\log}
=
-\frac32Q^2(QH+6MU).
\end{equation}
Using Eq.~\eqref{eq:bminusone},
\begin{equation}
b_u^{\log}
=
-\frac{3Q^3}{10}H
-\frac{18MQ^2}{5}U .
\label{eq:bulog}
\end{equation}

The gravitational logarithm follows from Eq.~\eqref{eq:Bgrav}. Defining
\begin{equation}
{\cal D}_h[y]
=
y''
-\frac{1}{f}
\left(
\frac{6}{r^2}
-\frac{4M}{r^3}
+\frac{2Q^2}{r^4}
\right)y ,
\end{equation}
one has
\begin{equation}
{\cal D}_h\left[
r^{-2}\log\left(\frac{r_0}{r}\right)
\right]
=
5r^{-4}+O(r^{-5}\log r).
\end{equation}
Since
\begin{equation}
\frac{4Q}{r^2}u^{(1)\prime}
=
-\frac{4Qb_{-1}}{r^4}+\cdots ,
\end{equation}
the resonant $r^{-4}$ equation is
\begin{equation}
5b_h^{\log}=-4Qb_{-1},
\end{equation}
and therefore
\begin{equation}
b_h^{\log}=\frac{6Q^3}{5}U .
\label{eq:bhlog}
\end{equation}
There is no term proportional to $H$ in $b_h^{\log}$.

Collecting Eqs.~\eqref{eq:bulog} and \eqref{eq:bhlog}, the logarithmic
part of the first-order correction is
\begin{equation}
\begin{pmatrix}
h_0^{(1)}\\[1mm]
u^{(1)}
\end{pmatrix}_{\log}
=
\frac{\log(r_0/r)}{r^2}
\begin{pmatrix}
0 & 6Q^3/5\\[1mm]
-3Q^3/10 & -18MQ^2/5
\end{pmatrix}
\begin{pmatrix}
H\\ U
\end{pmatrix}.
\end{equation}
Restoring the perturbative parameter $\gamma$ reproduces Eq.~\eqref{eq:rawlog}.
The four entries are therefore fixed locally by the asymptotic
recursion and do not depend on horizon matching. The arbitrary scale
$r_0$ shifts only the finite $r^{-2}$ coefficients.

Finally, transforming to the action-normalized fields derived in
Appendix~A gives
\begin{equation}
C_{\rm can}
=
C_{\rm raw}
\begin{pmatrix}
-4&0\\
0&1
\end{pmatrix}
=
\begin{pmatrix}
0&6Q^3/5\\
6Q^3/5&-18MQ^2/5
\end{pmatrix},
\end{equation}
which provides the canonical reciprocal matrix quoted in the main
text.


\begin{thebibliography}{99}

\bibitem{Flanagan:2007ix}
E. E. Flanagan and T. Hinderer,
``Constraining neutron star tidal Love numbers with gravitational wave detectors,''
Phys. Rev. D \textbf{77}, 021502 (2008).
\href{https://arxiv.org/abs/0709.1915}{arXiv:0709.1915}.

\bibitem{Binnington:2009bb}
T. Binnington and E. Poisson,
``Relativistic theory of tidal Love numbers,''
Phys. Rev. D \textbf{80}, 084018 (2009).
\href{https://arxiv.org/abs/0906.1366}{arXiv:0906.1366}.

\bibitem{Goldberger:2004jt}
W. D. Goldberger and I. Z. Rothstein,
``An effective field theory of gravity for extended objects,''
Phys. Rev. D \textbf{73}, 104029 (2006).
\href{https://arxiv.org/abs/hep-th/0409156}{arXiv:hep-th/0409156}.

\bibitem{Pereniguez:2021xcj}
D. Pere\~niguez and V. Cardoso,
``Love numbers and magnetic susceptibility of charged black holes,''
Phys. Rev. D \textbf{105}, 044026 (2022).
\href{https://arxiv.org/abs/2112.08400}{arXiv:2112.08400}.

\bibitem{Rai:2024lho}
M. Rai and L. Santoni,
``Ladder symmetries and Love numbers of Reissner-Nordstr\"om black holes,''
JHEP \textbf{07} (2024), 098.
\href{https://arxiv.org/abs/2404.06544}{arXiv:2404.06544}.

\bibitem{Barbosa:2025uau}
S. Barbosa, P. Brax, S. Fichet and L. de Souza,
``Running Love numbers and the effective field theory of gravity,''
JCAP \textbf{07} (2025), 071.
\href{https://arxiv.org/abs/2501.18684}{arXiv:2501.18684}.

\bibitem{Barbosa:2026qcv}
S. Barbosa, S. Fichet and L. de Souza,
``Running Love numbers of charged black holes,''
JCAP \textbf{09} (2026), 009.
\href{https://arxiv.org/abs/2602.00349}{arXiv:2602.00349}.

\bibitem{Noumi:2026shc}
T. Noumi and S. S. C. Wong,
``Extremal Love: tidal/electromagnetic deformability, logarithmic running and the weak gravity conjecture,''
JHEP \textbf{08} (2026), 033.
\href{https://arxiv.org/abs/2601.20962}{arXiv:2601.20962}.

\bibitem{Dunne:2004nc}
G. V. Dunne,
``Heisenberg-Euler effective Lagrangians: Basics and extensions,''
In From Fields to Strings: Circumnavigating Theoretical Physics, pp.~445-522 (World Scientific, 2004).
\href{https://arxiv.org/abs/hep-th/0406216}{arXiv:hep-th/0406216}.

\bibitem{Yajima:2000kw}
H. Yajima and T. Tamaki,
``Black hole solutions in Euler-Heisenberg theory,''
Phys. Rev. D \textbf{63}, 064007 (2001).
\href{https://arxiv.org/abs/gr-qc/0005016}{arXiv:gr-qc/0005016}.

%\cite{Ruffini:2013hia}
\bibitem{Ruffini:2013hia}
R.~Ruffini, Y.~B.~Wu and S.~S.~Xue,
``Einstein-Euler-Heisenberg Theory and charged black holes,''
Phys. Rev. D \textbf{88} (2013), 085004
%doi:10.1103/PhysRevD.88.085004
[arXiv:1307.4951 [hep-th]].
%123 citations counted in INSPIRE as of 25 Sep 2026

%\cite{Magos:2020ykt}
\bibitem{Magos:2020ykt}
D.~Magos and N.~Bret{\'o}n,
``Thermodynamics of the Euler-Heisenberg-AdS black hole,''
Phys. Rev. D \textbf{102} (2020) no.8, 084011
%doi:10.1103/PhysRevD.102.084011
[arXiv:2009.05904 [gr-qc]].
%95 citations counted in INSPIRE as of 25 Sep 2026


%\cite{Luo:2026srx}
\bibitem{Luo:2026srx}
H.~Luo, N.~Cao, X.~Y.~Chew, K.~G.~Lim, C.~Chen and D.~h.~Yeom,
``Purely Electric, Magnetic, and Dyonic Black Holes in Einstein-Euler-Heisenberg Theory,''
[arXiv:2607.21938 [gr-qc]].
%4 citations counted in INSPIRE as of 25 Sep 2026

%\cite{Luo:2026ndd}
\bibitem{Luo:2026ndd}
H.~Luo, N.~Cao, X.~Y.~Chew, K.~G.~Lim, C.~Chen and D.~h.~Yeom,
``The Analytical Solutions of Dyonic Black Holes in Einstein-Euler-Heisenberg Theory,''
[arXiv:2609.02380 [gr-qc]].
%1 citations counted in INSPIRE as of 25 Sep 2026




\bibitem{Breton:2021mju}
N. Bret\'on and L. A. L\'opez,
``Birefringence and quasinormal modes of the Einstein-Euler-Heisenberg black hole,''
Phys. Rev. D \textbf{104}, 024064 (2021).
\href{https://arxiv.org/abs/2105.12283}{arXiv:2105.12283}.


\bibitem{Lambiase:2024lvo}
G.~Lambiase, D.~J.~Gogoi, R.~C.~Pantig and A.~{\"O}vg{\"u}n,
``Shadow and quasinormal modes of the rotating Einstein{\textendash}Euler{\textendash}Heisenberg black holes,''
Phys. Dark Univ. \textbf{48}, 101886 (2025)
[arXiv:2406.18300 [gr-qc]].
%44 citations counted in INSPIRE as of 14 Sep 2026



%\cite{Verbin:2024ewl,Cimdiker:2026nmm,Lambiase:2024lvo,Gogoi:2026obd}
%\bibitem{Verbin:2024ewl}
%Y.~Verbin, B.~Pulice, A.~{\"O}vg{\"u}n and H.~Huang,
%``New black hole solutions of second and first order formulations of nonlinear electrodynamics,''
%Phys. Rev. D \textbf{111}, no.8, 084061 (2025)
%[arXiv:2412.20989 [gr-qc]].
%12 citations counted in INSPIRE as of 14 Sep 2026

%\cite{Cimdiker:2026nmm}
%\bibitem{Cimdiker:2026nmm}
%{\.I}.~{\.I}.~{\c{C}}imdiker, A.~{\"O}vg{\"u}n and Y.~Verbin,
%``Optical and orbital characterization of spherically symmetric static black holes of self-gravitating new nonlinear electrodynamics model,''
%Phys. Rev. D \textbf{114}, no.2, 024032 (2026)
%[arXiv:2603.10097 [gr-qc]].
%2 citations counted in INSPIRE as of 14 Sep 2026




%\cite{Gogoi:2026obd}
%\bibitem{Gogoi:2026obd}
%D.~J.~Gogoi, J.~Bora and A.~{\"O}vg{\"u}n,
%``Equatorial periodic orbits and gravitational wave signatures in Euler-Heisenberg black holes surrounded by perfect fluid dark matter,''
%JCAP \textbf{07}, 087 (2026)
%[arXiv:2604.11866 [gr-qc]].
%4 citations counted in INSPIRE as of 14 Sep 2026

%\cite{Becar:2026doz}
%\bibitem{Becar:2026doz}
%R.~Becar, P.~A.~Gonzalez, A.~Ovgun, J.~Saavedra and Y.~Vasquez,
%``Scattering, Hawking Radiation and Neutrino Energy Deposition in Euler-Heisenberg Black Holes Surrounded by Perfect Fluid Dark Matter,''
%[arXiv:2606.20931 [gr-qc]].
%3 citations counted in INSPIRE as of 25 Sep 2026





\bibitem{Moreno:2002gg}
C. Moreno and O. Sarbach,
``Stability properties of black holes in self-gravitating nonlinear electrodynamics,''
Phys. Rev. D \textbf{67}, 024028 (2003).
\href{https://arxiv.org/abs/gr-qc/0208090}{arXiv:gr-qc/0208090}.

%\cite{Chaverra:2016ttw}
\bibitem{Chaverra:2016ttw}
E.~Chaverra, J.~C.~Degollado, C.~Moreno and O.~Sarbach,
``Black holes in nonlinear electrodynamics: Quasinormal spectra and parity splitting,''
Phys. Rev. D \textbf{93} (2016) no.12, 123013
%doi:10.1103/PhysRevD.93.123013
[arXiv:1605.04003 [gr-qc]].
%31 citations counted in INSPIRE as of 25 Sep 2026



\bibitem{Daghigh:2021psm}
R. G. Daghigh and M. D. Green,
``Gravitational and electromagnetic radiation from an electrically charged black hole in general nonlinear electrodynamics,''
Phys. Rev. D \textbf{105}, 024055 (2022).
\href{https://arxiv.org/abs/2106.01412}{arXiv:2106.01412}.

\bibitem{Nomura:2021}
K. Nomura and D. Yoshida,
``Quasinormal modes of charged black holes with corrections from nonlinear electrodynamics,''
Phys. Rev. D \textbf{105}, 044006 (2022).
\href{https://arxiv.org/abs/2111.06273}{arXiv:2111.06273}.




\bibitem{Bastianelli:2008}
F. Bastianelli, J. M. D\'avila and C. Schubert,
``Gravitational corrections to the Euler-Heisenberg Lagrangian,''
JHEP \textbf{03} (2009), 086.
\href{https://arxiv.org/abs/0812.4849}{arXiv:0812.4849}.

\bibitem{Euler:1935zz}
H. Euler and B. Kockel,
``\"Uber die Streuung von Licht an Licht nach der Diracschen Theorie,''
Naturwissenschaften \textbf{23}, 246-247 (1935).

\bibitem{Heisenberg:1936nmg}
W. Heisenberg and H. Euler,
``Folgerungen aus der Diracschen Theorie des Positrons,''
Z. Phys. \textbf{98}, 714-732 (1936).
\href{https://arxiv.org/abs/physics/0605038}{arXiv:physics/0605038}.
English translation is the cited arXiv version.

\bibitem{Cardoso:2019upw}
V. Cardoso and F. Duque,
``Environmental effects in GW physics: tidal deformability of black holes immersed in matter,''
Phys. Rev. D \textbf{101}, 064028 (2020).
\href{https://arxiv.org/abs/1912.07616}{arXiv:1912.07616}.

\bibitem{Chakraborty:2024gcr}
S. Chakraborty, G. Comp\`ere and L. Machet,
``Tidal Love numbers and quasi-normal modes of the Schwarzschild-Hernquist black hole,''
(2024).
\href{https://arxiv.org/abs/2412.14831}{arXiv:2412.14831}.

\end{thebibliography}
\end{document}